\documentclass[epjST]{svjour}
\usepackage{graphics}
\usepackage{enumerate}

\begin{document}
\title{Can an interdisciplinary field contribute to one of the parent disciplines from which it emerged?}

\author{Anirban Chakraborti\inst{1}\fnmsep\thanks{\email{anirban@jnu.ac.in}} \and Dhruv Raina\inst{2}\fnmsep\thanks{\email{d\_raina@yahoo.com}} \and Kiran Sharma\inst{1}\fnmsep\thanks{\email{kiran34\_sit@jnu.ac.in}}}
\institute{School of Computational and Integrative Sciences, Jawaharlal Nehru University, New Delhi-110067, India \and Zakir Husain Centre for Educational Studies, School of Social Sciences, Jawaharlal Nehru University, New Delhi-110067, India}
\abstract{In the light of contemporary discussions of inter and trans disciplinarity, this paper approaches econophysics and sociophysics to seek a response to the question -- whether these interdisciplinary fields could contribute to physics and economics. Drawing upon the literature on history and philosophy of science, the paper argues that the two way traffic between physics and economics has a long history and this is likely to continue in the future.} 
\maketitle
\section{Introduction}
\label{intro}

\begin{quote}
\begin{flushleft}
\textit{``And how long do you think we can keep up this coming and going'' he asked.\\ Florentina Aziza had kept his answer ready...\\``Forever'' he said.}
\end{flushleft}

\begin{flushright}
-- Gabriel Garcia M\'arquez, \textit{Love in the Time of Cholera}

\end{flushright}
\end{quote}

In order to answer the question whether econophysics could contribute to physics and/or economics, we shall begin by charting out the emergence of interdisciplinarity, so as to comprehend the nature of exchange and circulation of concepts and metaphors between disciplines. It is these transactions that provide the conceptual bedrock for the emergence of interdisciplinarity. This is followed by a brief discussion on interdisciplinary practices and the subsequent unfolding of interdisciplinarity. And in the last section, we discuss the question just asked, which in a way can be reframed as: can an interdisciplinary field contribute to one of the parent disciplines from which it emerged? In other words, the question is about the reversal of the direction of conceptual influences. 

While the origins of interdisciplinary fields can be debated, two clarifications need to be made at the outset. In the first instance, we could take the view that we were always interdisciplinary in one way or another, but we could not be interdisciplinary before the disciplines concretized in stable institutional practices. And so, for this paper, at least the nineteenth century is the century of the stabilization of disciplines as recognized in the University of teaching and research \cite{wallerstein}. Secondly, while we could trace the origins of interdisciplinary fields to the early part of the 20th century, it is only in the second half of that century that the demand for and the discussion on interdisciplinarity became more cogent, coherent and urgent. One of the landmarks was the creation of research institutes and laboratories for solving military and strategic problems, interdisciplinary problem-focused research (or IDR), as it came to be called, and introduced university administrators to large scale collaborative projects on campus. The Manhattan project and operations research reinforced this instrumental discourse that Peter Weingart labeled pragmatic or opportunistic interdisciplinarity \cite{klein1}.

The spread of interdisciplinary research over the last half century is further signaled by the appearance of new social and cognitive forms that have altered the academic landscape, introducing new knowledge related practices that realign disciplinary relations. The term interdisciplinarity as Peter Weingart suggested, has become a metonym for innovation, and connotes everything from new ideas to product design \cite{klein2}. In the sciences, physics and biology or chemistry and biology instantiate the idea of interdisciplinarity for they provide the most striking examples of such sciences: biochemistry, molecular biology, materials sciences and now, of course, global networks to information science to chaos theory. It is during this period that we witness an enhanced cross-fertilization across several branches of physics from systems theory and sub-fields such as radioastronomy, biochemistry and plate tectonics. Since then philosophers of science have been attempting to characterize the different patterns of disciplinary relations. In his book, \textit{Integrating Scientific Disciplines}, Wiliam Bechtel sets out five types of interdisciplinary relations:
\begin{itemize}
  \item ``Developing conceptual links using a perspective in one discipline to modify a perspective in another discipline;
  \item Recognizing a new level of organization with its own processes in order to solve problems in existing fields;
  \item Using research techniques  developed in one discipline to elaborate a theoretical model in another;
  \item Modifying and extending a theoretical framework from one domain to apply in another;
  \item Developing a new theoretical framework that may conceptualize research in separate domains as it attempts to integrate them" \cite{bechtel1}.
\end{itemize}

The intensity of this boundary crossing led the National Research Council in the USA in the 1970s to declare that the emergence of new disciplines at the interfaces of physics and other sciences suggested that the boundary between physics and these other sciences was highly fluid and transient. Again in 1986, its report pointed out that all significant growth had occurred in what might be called \textit{interdisciplinary} borderlands, with blurred boundaries and a refocusing of scientific work towards problems characterized by unpredictability and complexity \cite{bechtel2} . The sociologist of science, Helga Nowotny, who co-authored \textit{The New Production of Knowledge}, and its sequel \textit{Re-Thinking Science. Knowledge and the Public in an Age of Uncertainty}, in `The Potential of Transdisciplinarity',  alerted us to a new conception of interdisciplinarity that is quite at odds with the 1960's conception. The earlier conception of interdiscipinarity was premised on the invention of a framework shared across disciplines to which each discipline contributed a bit. Transdisciplinarity is the next stage of interdisciplinarity that brings disciplines together in contexts, where new approaches arise out of their interaction – something not like a compound but a supercompound is produced. Three conditions have been seen to coexist in the formation of successful transdisciplinary formations:
 \begin{enumerate}
\item a common theoretical understanding, 
\item a mutual interpenetration of disciplinary epistemology,
\item a healthy disrespect for both disciplinary boundaries and institutional ones \cite{nowotny1}.
\end{enumerate} 
An interesting feature is that the terms  transdisciplinary and transdisciplinarity appear 14 times in Nowotny's paper while the terms transgressive or transgression appear almost half as many times and the author is aware of the relationship between the two, for it is pointed out explicitly that the prefix \textit{trans} is shared by the two sets of terms \cite{nowotny2}. This is a reflection of developments in two distinct directions. In the first case, is the recognition that what happens in disciplines is also happening in society, for example, the breakdown of functional differentiation between separate domains of social life, and the emergence of multitasking skills.  The second direction involves the co-evolution of society and knowledge in the sense that changing society becomes a factor in the production of knowledge. This is the context of application with the clarification that the context speaks back. Within the context of application then questions of the social implications of scientific knowledge must first be asked in the laboratory rather than when the knowledge comes out of the laboratory. This shift is essential if the knowledge produced is to be socially robust \cite{strathern1}.

The question increasingly asked within policy circles is, why the clamour is growing for intrerdisciplinarity, or the promotion of dialogue between disciplines? Nowotny's response is that the clamour for interdisciplinarity arises from the interests in generating better science. Since science is entering into a new social contract with society, it must respond to the call for an “openness on social issues and greater public accountability”. Clearly enough, interdisciplinary research can only be forged through collaboration. Some recent literature on the status of collaboration in the sciences suggests that there appears to be an erosion of the distinction between pure and applied research, both conceptually and in practice \cite{sheniderman}. 

In other words, interdisciplinarity even transgresses the compulsions of pragmatic, opportunistic or instrumental goals and may indeed entail a conceptual transfer that creates new theoretical alignments. The instrumental and conceptual triggers of interdisciplinarity bears resonances with those in the social sciences and humanities \cite{strathern1}. Nevertheless, there are significant differences between the two kinds of sciences. For one, in the human and social sciences unlike in the sciences, the instrumental and conceptual dovetail into one another. This distinction has been brought up because it is one thing to forge an interdisciplinarity between the different science disciplines and another thing between the sciences and the social sciences. In the former case, there is a family resemblance \footnote[1]{The term `family resemblances' is taken from Wittgenstein's \textit{Philosophical Investigations}. In this work he discusses a range of games, card games, ball games, etc., and invites us to see whether there is anything common to all the `proceedings' we call games. He then points out that: ``And the result of this examination is: we see a complicated network of similarities overlapping and criss-crossing: sometimes overall similarities". These similarities he calls family resemblances. We could as well extend this idea to the different mathematical games and techniques employed in a range of disciplinary fields \cite{Wittgenstein}.} between the sciences, while the entire literature on the human and social sciences marks the separation between the former and latter two. Yehuda Elkana once suggested the need for research students within the present matrix of research need to embrace risk and contradictions without being a Don Quixote \cite{Elkana}. The question then is, does the distinction really matter?

In the social and human sciences, the form of presentation of ideas moulds the perception of phenomena. But does this separate the two, for even Darwin borrows `notions of genealogy and descent' from the `language of human kinship'. However the natural sciences, eventually borrowings get `sedimented in the language of description'. So much so, after sometime little attention is paid to terms like ``killer cells", ``quarks", or ``dwarves" \cite{strathern2}. In the human sciences, it is imperative to pay attention to borrowings, the use of terms and concepts and their shifting connotation. In either case, new ideas and concept, appear in novel combinations, with practitioners knowing fully well that they could flow or migrate across disciplines. With meanings, boundaries also move and overlap, since ideas and techniques are constantly on the move, and it is researchers who transmit them in the process, creating a new division of labour directed towards novel ends \cite{Klein3}. 

The discussion on Darwinism  tells us something about conceptual flows between the natural and social sciences. The concept of evolution is often seen as a short hand for Darwinian evolution: `a process of descent with modification, whereby a species adapts to an environment through iterations of replication with random variation followed by natural selection' \cite{Gabora1}. Besides the term evolution does connote an unfolding, a narrative of how things change, a particular kind of change implying `the emergence of something from something else', `the idea of incremental and gradual change', in opposition to revolutionary break \cite{Ridley1}. For over a century and a half, the term, ``evolution" has acquired currency across the biological and social sciences, alluding to dynamical changes of state without implying natural selection. In other words, natural selection is only one part of the process of evolution. If Darwinism is the special theory of evolution, Ridley  suggests that there should be a general theory of evolution that extends biology to society, the economy, to technology, to language, history, education, knowledge, morality. `The general theory of evolution says that things do not stay the same: they change gradually but inexorably: they show path dependence', descent with modification, trial and error and selective persistence \cite{Ridley2}. This recognition has prompted several recent endeavours to formulate a general theory of evolution, one version of the theory being premised on the idea that `all entities evolve through a reiterated process of interaction with a context', there being different forms of evolution depending on the degree of non-determinism, the degree of contextuality and the degree of context-driven change \cite{Gabora2}. We can well infer that the history of interdisciplinary transactions between the natural sciences and the so called social sciences itself is more than two hundred years old, and much of this has not been of the nature of purely pragmatic or instrumental transaction but has operated at a deeper conceptual level, involving conceptual borrowing and subsequent naturalization in the new disciplinary or interdisciplinary field. 

\section{The grammar of interdisciplinary formation}

The naming of an interdisciplinary field provides some insight into the process of formation. As an illustration, consider the formation of interdisciplinary fields such as biophysics and biochemistry -- sociophysics and econophysics -- the suffix of the interdisciplinary field signals the discipline from which the framework, method, protocol is inherited and projected onto another field. The procedure of nomenclature reveals the objectives, and protocols of disciplinary formation and the modes of entry into the discipline. However, there could be fields, which are combinations of several disciplines in which the disciplinary suffix is still the one that defines the new formation-- molecular biophysics.

 There could be fields, which though labelled interdisciplinary do not obey such a law of formation. The case of material sciences, ecology and many others stand out: here it could be reckoned that the modes of formation are quite different -- less mathematically formalized, but where tacit knowledge plays an important role, and where contending frameworks are juxtaposed (natural history versus thermodynamics; solid state chemistry versus metallurgy), etc.

But the important point worth noting is that it is easier to construct interdisciplinarity amongst the sciences due to greater `family resemblances'. This paper does not take the distinction between the sciences and social sciences to be a facile one. Drawing upon Philip Mirowski's work, it could be argued that the two way traffic between physics and economics has a long history \cite{Mirowski1}. But as and when that happens, it is the methodological expansion of the sciences into the social sciences that we see -- very often reflected in a second wave of mathematisation, that we could label as the mathematisation of the social sciences that in turn is contiguous with the legacy of the social sciences. In this second wave, we could include the extension of methods of statistical mechanics, graph theory and network theory to study  economics, social and trade networks, social institutions, etc. The development of both sociophysics and econophysics can be framed within the latter rubric.

Adopting a Kuhnian language, the extension of the sciences into the social sciences, in this case, entails the adoption of an analogy or mathematical technique(s) from physics (or the possible application of graph theory or network theory) into the social domain. The conceptual analogy operates as kind of Kuhnian exemplar; but as with analogies-- these are not one to one; rather are circumscribed by constraints, diversity and limits of applicability. These factors provide the envelope that forms the boundary of the new interdisciplinary formation, while respecting the contours of the parent disciplines. More than anything else this serves as a mechanism of disciplinary conflict resolution.

\section{Between `Value` and `Energy`}

Returning to econophysics and sociophysics; are based on the extension of statistical mechanics, graph theory and network theory to the social domain. Methodologically, this entails a mathematical reduction of social entities, denuding them of the phenomenological dimension of human experience, or the socialization of political conflict and tension. This is done through the introduction/ fabrication of bipolar/ dichotomous entities that stretch distinctions distributed over a spectrum into binaries. Secondly, the suppression of the phenomenological translates the modelling into the realm of statistical distributions, where these subjectivities are reduced to arithmomorphic entities subject to the logic of probabilities. It would be interesting then to see, where these interdisciplinary fields finally migrate to and settle. Taking our example of the first type of  interdisciplinarity, the interdisciplinary field most likely lodges itself as a sub-field of the prefixed parent discipline, where biophysics and biomathematics are lodged in schools of the biological sciences. In the second type mentioned where an assemblage of fields comes together, the possibility of the emergence of an autonomous interdisciplinary field becomes increasingly likely. There is a third kind of interdisciplinarity one could envisage. The emergence of a generic mathematical technique such as nonlinear mathematics, chaos theory or fractals, could trigger of a number of interdisciplinary endeavors; for example, atmospheric and ocean sciences. These generic methods could themselves emerge within the womb of a formation; now institutionally structured under the rubric of computational and integrative sciences.

As a distinct exercise, it would be interesting to historically trace the invention and migration of interdisciplinary fields -- which of them became sub-disciplines within larger disciplinary frames and which of them acquired an interdisciplinary autonomy. The historical study could itself provide insights into the temporal evolution of the field.

The evolution of an interdisciplinary field can also be interpreted in terms of the circulation of a metaphor or a concept that originates in a discipline or starts out from a discourse situated in time, that is taken up in another discipline and gradually is naturalised, as it acquires a conceptual life. Conservation principles in physics could be said to belong to such a family \cite{Mirowski2}. Discussing disciplinary histories, Mirowski argues that there is an uncanny resemblance between neoclassical economics and physical theories. The argument goes that from the 1870's neoclassical economists framed their arguments and models on lines similar to those of physical theories. Central concepts that circulated through physical and social theory, travelling to and fro, were ``energy" in physics and ``utility" in economics.
The kinetic exchange models in physics are stochastic models which have had a straightforward interpretation in terms of energy exchanges in a gas. However, starting from 2000, they have been shown to be suitably adapted and used to study problems in the social sciences \cite{Chakraborti1,Chatterjee,Lallouache,Patriarca,Chakraborti2,Kiran} . Only recently has  it been shown that the dynamics of the kinetic exchange models could also be derived from standard microeconomic theory, although such dynamics are often criticized for being based on an approach that is far from an actual economics perspective \cite{Chakrabarti}. The standard economic theory ordinarily assumes that the activities of the individual agents are driven by the ``utility maximization'' principle. In the kinetic wealth exchange models, the economic agents trade ``wealth'' that his considered analogous to physical  particles exchanging ``energy". The entropy maximization (or energy minimization) principle valid for the physical kinetic theory of gases, has thus a one-to-one correspondence with the utility maximization for all the economic agents in a minimal model of an economy. Such qualitative analogies and metaphors have circulated between the two disciplines and both physicists and mathematicians have observed this, in the case of maximization principle. The underlying idea is that a large number of molecules and a large social group have many common features, including the fact that they may be predictable due to the high number of their components despite their intrinsic random character. Boltzmann wrote that ``The molecules are like individuals, . . . and the properties of gases only remain unaltered, because the number of these molecules, which on the average have a given state, is constant'', when writing about the foundations of statistical mechanics \cite{Ball,Boltzmann}.

In his book \textit{Populare Schriften}, Boltzmann (1905) advocates and appraises the systematic development of statistical mechanics by Josiah Willard Gibbs, 
\begin{quote}
``This opens a broad perspective, if we do not only think of mechanical objects. Let’s consider to apply this method to the statistics of living beings, society, sociology and so forth."
\end{quote}

Ettore Majorana, an Italian theoretical physicist who worked on neutrino masses and disappeared suddenly under mysterious circumstances on a trip from Palermo to Naples at the young age of 32, also had a remarkable view on the relation between the sciences and social sciences. An article was found by Majorana's brother among his files, which was then published in 1942, in the international Italian journal \textit{Scientia} \cite{Majorana}. The article  presented  the  physicist's perspective about the value of statistical laws in physics and social sciences to scholars of sociology and economics, and also considered the philosophical aspects related  to  the  nature  and  value  of  deterministic  and  statistical  laws  in  science.  In  the  article,  Majorana considered quantum mechanics as an irreducible statistical theory, since the theory was unable to describe the time evolution of a single particle or atom in a controlled environment at a deterministic level. Majorana therefore concluded that quantum mechanics suggests that there was an ``essential analogy  between  physics  and  the  social  sciences,  between  which  an  identity  of  value  and  method  has  turned  out''.

Another famous scientist, Frederick Soddy, noted for his work on radioactivity, had published the book entitled ``Wealth, Virtual Wealth and Debt'', where he proposed that real wealth is derived from the energy use in transforming raw materials into goods and services, and not from monetary transactions, thereby again connecting the sciences and the social sciences. Two Indian physicists, Meghnad Saha and B. N. Srivastava, in their textbook entitled \textit{A Treatise on Heat} (1931) had used the example of reconstructing a distribution curve for incomes of individuals in a country to illustrate the problem of determining the distribution of molecular velocities in kinetic theory.

It is also worth mentioning that many sociologists and economists were trained in sciences and engineering. Vilfredo Pareto possessed a degree in mathematical sciences and a doctorate in engineering, and while working as a civil engineer, he had collected statistics that demonstrated in 1897 that the distributions of income and wealth in a society follows a power law, known today as the Pareto law \cite{Pareto}. The influential American economist Irving Fisher was a student of the physicist J.W. Gibbs.  The first Nobel Prize winner in economics was Jan Tinbergen,
a physicist by training (having completed his Ph.D. from the University of Leiden in 1929 on ``Minimisation problems in Physics and Economics'', under the supervision of Ehrenfest).

A number of mathematical concepts and formalism  traveled from physics to economics particularly from Newtonian mechanics and classical thermodynamics \cite{Mirowski2,Smith}, which culminated in the neoclassical concept of mechanistic ``equilibrium'' where the ``forces'' of ``supply and demand'' balanced each other.  
While performing analogous conceptual functions in the respective disciplines, there were several similarities in the mathematical formalisms \cite{invisiblehand}. This is probably why, and not without reason, that Mirowski finds it difficult to sustain the distinction between the human and social sciences on the one hand and the natural sciences on the other \cite{Mirowski3}. Drawing on the work of the philosopher of science, Emile Meyerson, it is suggested that not only was there a circulation from physics to economics but prior to there was one from economics to physics –- most likely through the discussion on equivalences and values. Furthermore, the concept of energy itself, which within the discussion on econophysics provides one of the links for forging an operational analogy between physics and the economy, in the nineteenth century, was certainly not the property of physics, for in both economics and physics the ideas of `body, motion and value' played a central role in reinforcing the validity of conservation principles.  This was seen as a way of reconciling the `flux and diversity of experience' with laws and reveals the ways in which ideas from the economy influence science \cite{Mirowski4}.   

\section{Concluding remarks}

In the light of the above discussions, insurmountable historical evidence suggests that there is a circulation of metaphors from economics to physics and later of methods and mathematical formalisms from physics to economics in the nineteenth century. So there is no reason to believe that this mutual enrichment is likely to be arrested. In the particular case discussed here, econophysics is an interdisciplinary outcome of this long mutually beneficial relationship. As and when econophysics or sociophysics generate metaphors, techniques and formalisms of their own – in whatever sense, there is no \textit{a priori} impediment that prevents them from travelling in the reverse direction -– that is to physics or economics, since the `family resemblance' between the three provides at least one condition of commonality. The argument can be pushed further to qualify an earlier claim that it is easier to construct an interdisciplinary field between the different natural sciences than between the natural and social sciences. Econophysics provides us with a promising counter example of an interdisciplinary field forged across the postulated divide.

%

%
%

\end{document}